# Electrical manipulation of spin states in a single electrostatically gated transition-metal complex

Edgar A. Osorio,[1] Kasper Moth-Poulsen,[2] Herre S.J. van der Zant,[1*] Jens Paaske,[2] Per Hedegård,[2] Karsten Flensberg,[2] Jesper Bendix[2] and Thomas Bjørnholm[2]

[1] Kavli Institute of Nanoscience, Delft University of Technology, PO Box 5046, 2600 GA, The Netherlands

[2] Nano-Science Center (Department of Chemistry and Niels Bohr Institute), University of Copenhagen, Universitetsparken 5, DK-2100, Copenhagen, Denmark.

[*]EMAIL: h.s.j.vanderzant@tudelft.nl

ABSTRACT. We demonstrate an electrically controlled high-spin (S=5/2) to low-spin (S=1/2) transition in a three-terminal device incorporating a single $Mn^{2+}$ ion coordinated by two terpyridine ligands. By adjusting the gate-voltage we reduce the terpyridine moiety and thereby strengthen the ligand-field on the Mn-atom. Adding a single electron thus stabilizes the low-spin configuration and the corresponding sequential tunnelling current is suppressed by spin-blockade. From low-temperature inelastic cotunneling spectroscopy, we infer the magnetic excitation spectrum of the molecule and uncover also a strongly gate-dependent singlet-triplet splitting on the low-spin side. The measured bias-spectroscopy is shown to be consistent with an exact diagonalization of the Mn-complex, and an interpretation of the data is given in terms of a simplified effective model.

The vision for single-molecule electronics, and spintronics applications, is to tailor the electrical and magnetic properties of a device already in the chemical synthesis of the molecule. This vision holds the promise of unprecedented functionalization of the device, in terms of built-in mechanical, conformational, optical and magnetic properties of the isolated molecule. A molecule bridging two electrodes does not, however, preserve all the well-characterized properties of the isolated molecule. This is due to the strong influence of the nearby electrodes [1-5], which at the same time allows manipulations of the molecule which are note possible by chemistry. In this letter, we report on charge transport at low-temperature in three-terminal molecular junctions containing a *Mn*-transition metal complex [6]. Leaving aside their vast importance in bioinorganic chemistry, transition metal complexes [7-8] are particularly interesting for molecular spintronics devices [9-12]. With the present findings, we demonstrate direct electrical control of the spin-ground state of a single metal complex, and show that the gate electrode in these experiments can be conceived as an electrically tunable external ligand.

Our devices are made by electromigration [13] of a gold wire in a solution of the molecules, using a feedback mechanism [14] combined with self-breaking [15] (see Fig. 1a for a likely structure of the device). Details of our fabrication and junction preparation are described in Ref. [16] and the molecule synthesis and characterization is described in section S1 of the supplementary information. We focus on a particular $Mn^{2+}$ junction with intermediate coupling to source and/or drain electrodes, and concomitant Kondo effects [17-19] as well as pronounced inelastic cotunneling lines [20-21]. The transport measurements are represented in Fig. 2, in which we plot the differential conductance as a function of the applied bias, and gate voltage, showing the typical 'diamond' signatures of Coulomb blockade (CB). As we shall argue (cf. also section S2 of the supplementary information), these data reflect two molecules in parallel: one (molecule B) giving rise to the two sharp white crosses separating regions (I-III) and another (molecule A) leading to the much broader cross separating regions (1) and (2). In this letter, we focus on the more strongly coupled molecule (A), which exhibits a pronounced blocking of low-bias transport at the charge-degeneracy point. This molecule (A) also gives rise to sharp inelastic cotunneling lines, observed as faint edges within the black background of Fig. 2b, pervading all of region (2) and revealing a marked gate dependence of the spin-excitations on this molecule.

In order to resolve the spin-configurations in the two charge-states (1-2) of molecule A, we have measured magnetic field dependence of the inelastic cotunneling lines (Fig. 3). In a field of 10 Tesla, the inelastic cotunneling lines in region (2) are observed to cross at zero bias voltage for a gate voltage $V_C$ denoted by an orange dot in Fig. 3b. On the left hand side of this crossing, three equally spaced lines are observed (red, blue, and green lines). On the right hand side, blue, and green excitations become gate independent. This behaviour with magnetic field points at an interesting interplay between a singlet (S), and a triplet (T) state [22-23] with a gate-dependent antiferromagnetic exchange coupling, $J(V_G)$, as illustrated in Fig. 3c (bottom panel). Considering the Zeeman splitting of the triplet states (Fig. 3d: bottom panel), the excitation pattern observed in charge state (2) indicates that the crossing point at zero bias voltage marks a transition from a singlet ground state (left) to a triplet ground state (right). On the left hand side of $V_C$ the excitations from the singlet ground state to the three components of the triplet give rise to three equally spaced resonances [21,23] separated by $g\mu_B B$. The splittings in Fig. 3b yield a *g*-factor of g ≈ 1.8. On the right hand side, the $T_{-1}$ component of the triplet is the ground state, and the observed resonances are consistent with inelastic cotunneling from the $T_{-1}$ state to the singlet S, $T_0$ and $T_{+1}$ states [21,23]. Again, the energy diagram in Fig. 3d predicts the $T_{-1} \rightarrow S$ transition to be gate dependent (red line), whereas for the $T_{-1} \rightarrow T_0$ and the $T_{-1} \rightarrow T_{+1}$ transitions (blue and green lines) no gate dependence is expected. The fact that $V_C$ moves towards smaller gate voltages for increasing magnetic field is also consistent with the energy diagram presented in Fig. 3d. Further details on the assignment of spin-states is provided in the supplementary information, section S3. In the left-most section of (1) and right-most section of (2), we observe zero-bias Kondo-peaks indicating spin-degenerate ground states. As shown in Fig. 3f, these undergo a simple Zeeman splitting in a magnetic field with corresponding *g*-factors of 2.1 ± 0.3, and 1.9 ± 0.3, respectively (cf. supp.info. section S5). As shown below, this detailed magnetic field dependence of the bias-spectroscopy allows us to build up a consistent model for the electronic configurations of molecule A in the measured gate-range.

From our bulk SQUID susceptibility measurements (cf. supp. info. section S1) we find that the metal complex in its crystalline form has a high-spin (HS) ground state with *S=5/2*. This is as expected from ligand-field theory since exchange interactions at the metal core are large enough [24, 25] to overcome the nearly octahedral energy splitting $\Delta_{oct}$ [26]. Thus it would be natural to assign this high-spin ground state with *N=5* *d*-electrons on the *Mn*-atom to region (1), which makes region (2) a 6-particle state, with an extra electron fluctuating between the *Mn* *d*-orbitals and a ligand state. From the magnetic field dependence of the bias-spectroscopy we know that regions (2,I-II) have a singlet ground-state and therefore the *Mn*-centre must have changed into a low-spin (LS) configuration, which can be paired with the single spin 1/2 of an added electron to produce the observed singlet. Several factors could contribute to stabilizing the unusual [27] low-spin state of the manganese(II) centre including loss of solvation and counter ion interactions, but the likely cause is the increase in ligand-field strength of the terpyridine ligand system upon reduction. It should be noted that reduction of terpyridine ligands coordinated to divalent metal centers is known, even for systems in solution [28]. Thus, charging the ligands by increasing the gate-voltage can indeed be expected to inflict a transition from HS to LS in the *Mn*-centre.

This change in spin configuration leads to a suppression of the current which would otherwise flow at the charge degeneracy point. Due to the high-spin *Mn*-centre, the addition of a single electron cannot give rise to a spin-singlet. Transport at low bias is suppressed since the two neighbouring ground states, (N=5, S=5/2) and (N=6, S=0), are not coupled via one-electron fluctuations. Sequential tunnelling near the *N*=5,6 charge degeneracy point is therefore restricted to excited states, which can be reached only above a finite bias corresponding roughly to the nearby inelastic cotunneling threshold. This spin-blockade [29] is thus lifted when the bias-voltage is large enough to populate the excited states, i.e. states with *N=5, S=1/2* in region (1) and *N=6, S=1* in region (2). Unlike the respective ground-states, these low-lying excited states do couple via one-electron fluctuations and transport takes place by means of so-called cotunneling assisted sequential tunnelling, known from experiments on quantum dots [30] (cf. section S7 of supplementary information). Finally, we note that the observation of zero-bias Kondo peaks in the left part of region (1) and the right part of region (2) is again consistent with the *Mn*-centre having *S=5/2* in (1) and *S=1* in the right part of (2).

To gain further insight about the electronic configurations for the two charge-states of molecule A, we have diagonalized the fully interacting 5 and 6-electron problems for an isolated *Mn*-complex. We include the three $t_{2g}$, and two $e_g$ d-orbitals on the Mn-atom, split by an energy $\Delta$ due to the nearly octahedral ligand-field, together with two ligand states made out of $p_z$-orbitals on the N-atoms. We include electrostatic Coulomb repulsion, *U*, and ferromagnetic exchange, *K*, among the *d*-electrons. Furthermore, electrons on either of the ligand orbitals are allowed to tunnel onto a corresponding $t_{2g}$ orbital with matching symmetry (cf. supp.info. section S6). Interestingly, we find that for a large range of realistic parameters the calculations reproduce the main features of the data: A *high-spin* (S=5/2) 5-electron ground state for low gate voltages and a low-spin (S=0) 6-electron ground state for larger gate voltages. Furthermore, the model exhibits low-lying (*meV* range) excited states having respectively spin 1/2 (N=6) and spin 1 (N=5), needed to explain the cotunneling assisted sequential tunnelling setting in at finite bias above the spin-blocked N=5,6 charge-degeneracy point. The singlet character of the 6-electron ground state requires the *Mn*-atom to be stabilized in a low-spin configuration when adding an electron to the neutral complex, indicating a charging induced increase in the ligand-field splitting Δ. This sudden increase we ascribe to the intra-ligand Coulomb repulsion with the extra electron, which will move the occupied ligand levels upwards and closer into resonance with the $e_g$ levels.

The gate dependence of the singlet-triplet splitting observed throughout region (2, I-III) can readily be explained by a difference in gate-coupling for the two terpyridine moieties, arising naturally in the asymmetric device configuration envisioned in Fig. 1a. This asymmetric configuration is consistent with the broad, yet relatively low, sequential tunnelling conductance ridge separating regions (1) and (2) in Fig. 2a, which is to be expected with a large difference in tunnel-couplings to source, and drain electrodes. Additional confirmation of asymmetric coupling is presented in section S4 of the supplementary information in which we report the observation of a faint, nearly vertical, line pinned to the singlet-triplet transition point. Calculations show that this line is visible only with a substantial source-drain asymmetry in the tunnel couplings. In an asymmetric geometry, electrons on the terpyridine closer to an electrode or on the *Mn*-atom are screened by the nearby metallic electrode and will therefore be much weaker coupled to the back-gate potential than electrons on the central terpyridine moiety. Figure 4 illustrates the results of our exact diagonalization in terms of the few many-body states which dominate the exact result. From these states it becomes apparent that the triplet state gains an extra ferromagnetic exchange energy compared to the singlet state when an electron is shifted more toward the central (gate sensitive) ligand orbital by further increasing $V_g$ and moving right in region (2,I-III).

**Acknowledgment.** We thank Kevin O'Neill for the work on the electromigration, Maarten Wegewijs for discussions, and Titoo Jain for graphical assistance. Financial support is obtained from the Dutch Organization for Fundamental Research on Matter (FOM), the 'Nederlandse Organisatie voor Wetenschappelijk Onderzoek' (NWO), the EU FP7 project SINGLE (contract#: 213609), and the Danish Agency for Science, Technology and Innovation (JP).

**Supporting Information Available**: Details on the molecule synthesis and characterization are given in section S1. Section S2 discusses the assignment of the low-bias features for the two molecules in parallel. Section S3 gives additional details to support the assignment of the different spin states as discussed in the main text. Section S4 explains the appearance of a sharp resonance line pinned to the singlet-triplet degeneracy as due to enhanced cotunneling. Section S5 shows the temperature and magnetic field dependence of the various Kondo resonances observed in the stability plot. Section S6 discusses model calculations carried out in order to understand how shifts of the ligand orbital energies are required to account for the observations, in particular the low-spin states. Section S7 deals with calculations carried out in order to understand the spin blockade phenomena observed in the experiments within the framework of the so-called *cotunneling assisted sequential tunneling*. This material is available free of charge via the Internet at http://pubs.acs.org.

**References**

(1) Osorio, E. A.; O'Neill, K.; Stuhr-Hansen, N.; Nielsen O. F.; Bjørnholm, T.; van der Zant, H. S. J. *Adv. Mater.* **2007**, *19*, 281-285.

(2) Osorio, E. A.; O'Neill, K.; Wegewijs, M.; Stuhr-Hansen, N.; Paaske, J.; Bjørnholm, T.; van der Zant, H. S. J. *Nano Lett.* **2007**, *7*, 3336.

(3) Yeganeh, S.; Galperin, M.; Ratner, M. A. *J. Am. Chem. Soc.* **2007**, *129*, 13313.

(4) Kaasbjerg, K.; Flensberg, K. *Nano Lett.* **2008**, *8*, 3809-3814.

(5) Moth-Poulsen, K.; Bjørnholm, T. *Nat. Nanotechnology.* **2009**, *24*, 551-556.

(6) Moth-Poulsen, K.; Bendix, J.; Hammershøj, P.; and Bjørnholm, T. *Acta Cryst*. **2009**, *C65*, 14-16.

(7) a) Albrecht, T.; Moth-Poulsen, K.; Christensen, J. B.; Guckian, A.; Bjørnholm, T.; Vos, J. G.; Ulstrup, J. *Faraday Discuss*. **2005**, *131*, 265. b) Albrecht, T.; Guckian, A.; Ulstrup, J.; Vos, J. G. *Nano Lett*. **2005**, *5*, 1451–1455.

(8) Rajadurai, C.; Schramm, F.; Brink, S.; Fuhr, O.; Ghafari, M.; Kruk, R.; Ruben, M. *Inorg. Chem*. **2006**, *45*, 10019-10021.

(9) Bogani, L.; Wernsdorfer, W. *Nature Mater*. **2008**, *7*, 179-186.

(10) Sanvito, S.; Rocha, A. R. *J. Comput. Theor. Nanosci*. **2006**, *3*, 624–642 (2006).

(11) Lehmann, J.; Gaita-Ariño, A.; Coronado, E.; Loss, D. *Nature Nanotechnology*, **2007**, *2*, 312-317.

(12) Ruben, M.; Landa, A.; Lörtscher, E.; Riel, H.; Mayor, M.; Görls, H.; Weber, H. B.; Arnold, A.; Evers, F. *Small,* **2008**, *4*, 2229-2235.

(13) Park, H.; Lim, A. K. L.; Alivisatos, A. P.; Park, J.; McEuen, P. L. *Appl. Phys. Lett*. **1999**, *75*, 301- 303.

(14) Strachan, D. R.; Smith, D. E.; Johnston, D. E.; Park, T. H.; Therien, M. J.; Bonnell, D. A.; Johnson, A. T. *Appl. Phys. Lett*. **2005**, *86*, 043109.

(15) O'Neill, K.; Osorio, E. A.; van der Zant, H. S. J. *Appl. Phys. Lett*. **2007**, *90*, 133109.

(16) Osorio, E. A.; Bjørnholm, T.; Lehn, J. M.; Ruben, M.; van der Zant, H. S. J. *J. Phys.: Condens. Matter*. **2008**, *20*, 374121.

(17) Goldhaber-Gordon, D.; Shtrikman, H.; Mahalu, D.; Abusch-Magder, D.; Meirav, U.; Kastner, M. A. *Nature*. **1998**, *391*, 156–159.

(18) Cronenwett, S. M.; Oosterkamp, T. H.; Kouwenhoven, L. P. *Science*, **1998**, *281*, 540–544.

(19) Nygård, J.; Cobden, D. H.; Lindelof, P. E. *Nature*. **2000**, *408*, 342-346.

(20) De Franceschi, S.; Sasaki, S.; Elzerman, J. M.; van der Wiel, W. G.; Tarucha, S.; Kouwenhoven, L. P. *Phys. Rev. Lett*. **2001**, *86*, 878-881.

(21) Paaske, J.; Rosch, A.; Wölfle, P.; Mason, N.; Marcus, C. M.; Nygård, J. *Nat. Phys*. **2006**, *2*, 460-464.

(22) Kogan, A.; Granger, G.; Kastner, M. A.; Goldhaber-Gordon, D.; Shtrikman, H. *Phys. Rev. B.* **2003**, *67*, 113309.

(23) Roch, N.; Florens, S.; Bouchiat, V.; Wernsdorfer, W.; Balestro, F. *Nature.* **2008**, *453*, 633-637.

(24) Borel, A.; Daul, C. A. Theochem. **2006**, *762*, 93-107.

(25) Brorson, M.; Schäffer, C. E. *Inorg. Chem.* **1988**, *27*, 2522.

(26) Lever, A. B. P. Studies on Physical and Theoretical Chemistry: Inorganic Electronic Spectroscopy (Second Edition); Elsevier: Amsterdam, **1984**.

(27) For examples of rare low-spin manganese(II) complexes see: Bendix, J.; Weyhermüller, T.; Bill, E.; Wieghardt, K. *Angew. Chem. Int. Ed.* **1999**, *38*, 2766-2768 and references therein.

(28) Braterman, P. S.; Song, J.-I.; Peacock, R. D. *Inorg. Chem.* **1992**, *31*, 555.

(29) Weinmann, D.; Hausler, H.; Kramer, B. *Phys. Rev. Lett.* **1995**, *74*, 984-987.

(30) Schleser, R.; Ihn, T.; Ruh, E.; Ensslin, K.; Tews, M.; Pfannkuche, D.; Driscoll, D. C.; Gossard, A. C. *Phys. Rev. Lett.* **2005**, *94*, 206805.

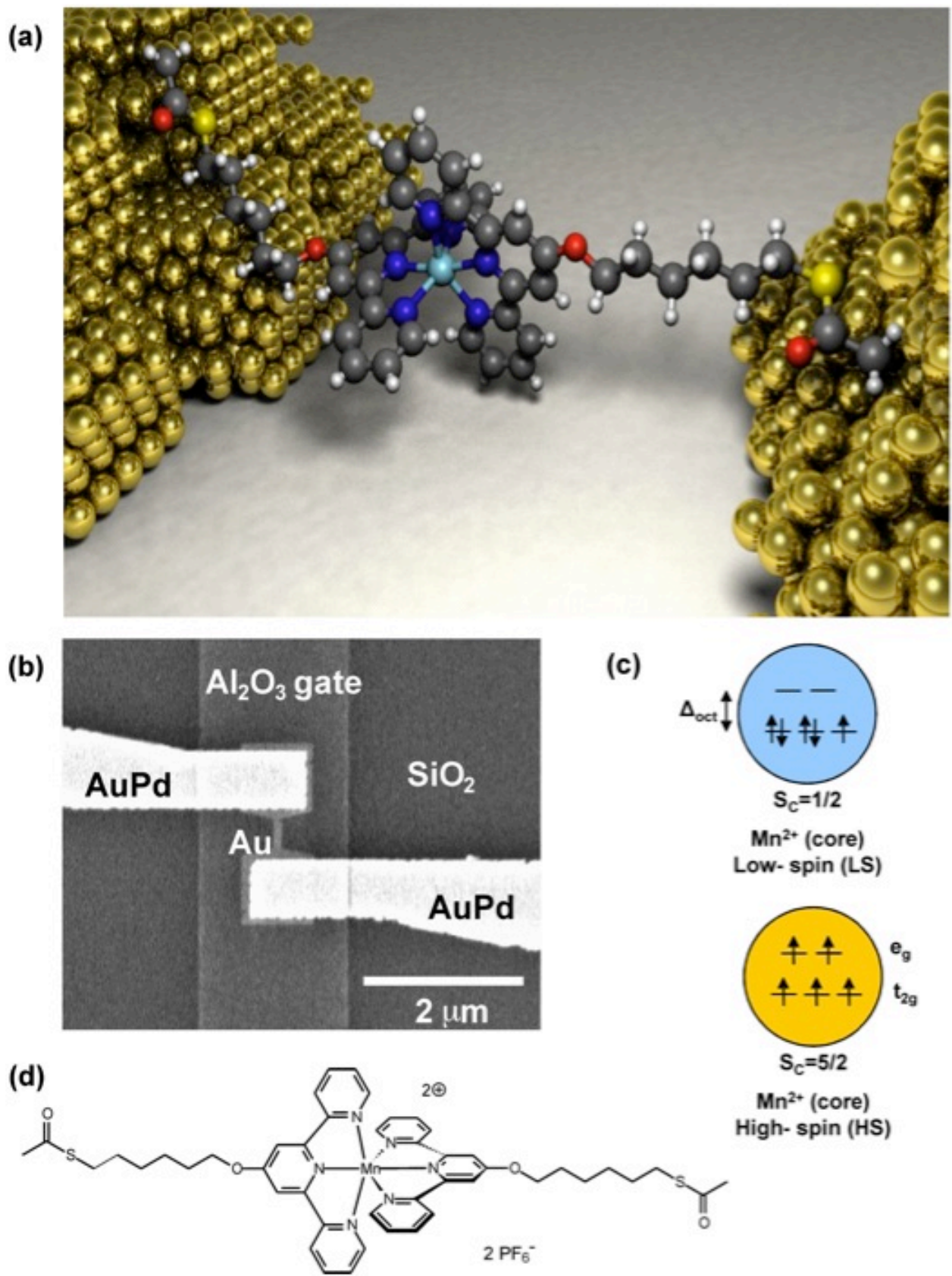

**Figure 1.** (a) Schematic device lay-out and artistic impression of a ([Mn(terpy-O-(CH$_2$)$_6$-SAc)$_2$)]$^{2+}$) molecule bonded to two gold electrodes. The asymmetric geometry illustrates a likely realization of the device which gives rise to asymmetric coupling to source, and drain electrodes, and the difference in gate-coupling to the two ligand moieties which is implied by our transport data. (b) Fabricated device prior to breaking the small gold wire in the middle by electromigration. The junction is fabricated on top of an aluminium gate electrode, which is oxidized in air to form a 2 to 4 nm thick $Al_2O_3$ layer and at low temperatures, substantial leakage currents are typically observed for voltage above ± 4 V. Bridges are electromigrated in the molecule solution at room temperature by ramping a voltage until a decrease in the conductance is observed, upon which the applied voltage is returned to 100mV; the cycle is repeated until a target resistance of 5kΩ has been reached. The electromigrated bridges are then left in the molecule solution for about one hour to allow for molecular self-assembly and "self-breaking" of the constricted gold wire. Last, the sample space is evacuated and the cooling procedure to 1.7 K starts. (c) Two different $d^5$ electronic configurations of the $Mn^{2+}$ core with respectively low, and high spin are given. The *d*-orbitals on the *Mn* atom are split by the nearly octrahedral ligand field of the organic terpyridine cage into three (lower) $t_{2g}$ orbitals and two (upper) $e_g$ orbitals. (d) Molecular structure of ($[Mn(terpy-O-(CH_2)_6-SAc)_2)]^{2+}$). The derivative has $CH_6$ alkane chains attached to the ligands and acetyl protected thiol end groups to ensure bonding with the gold electrodes.

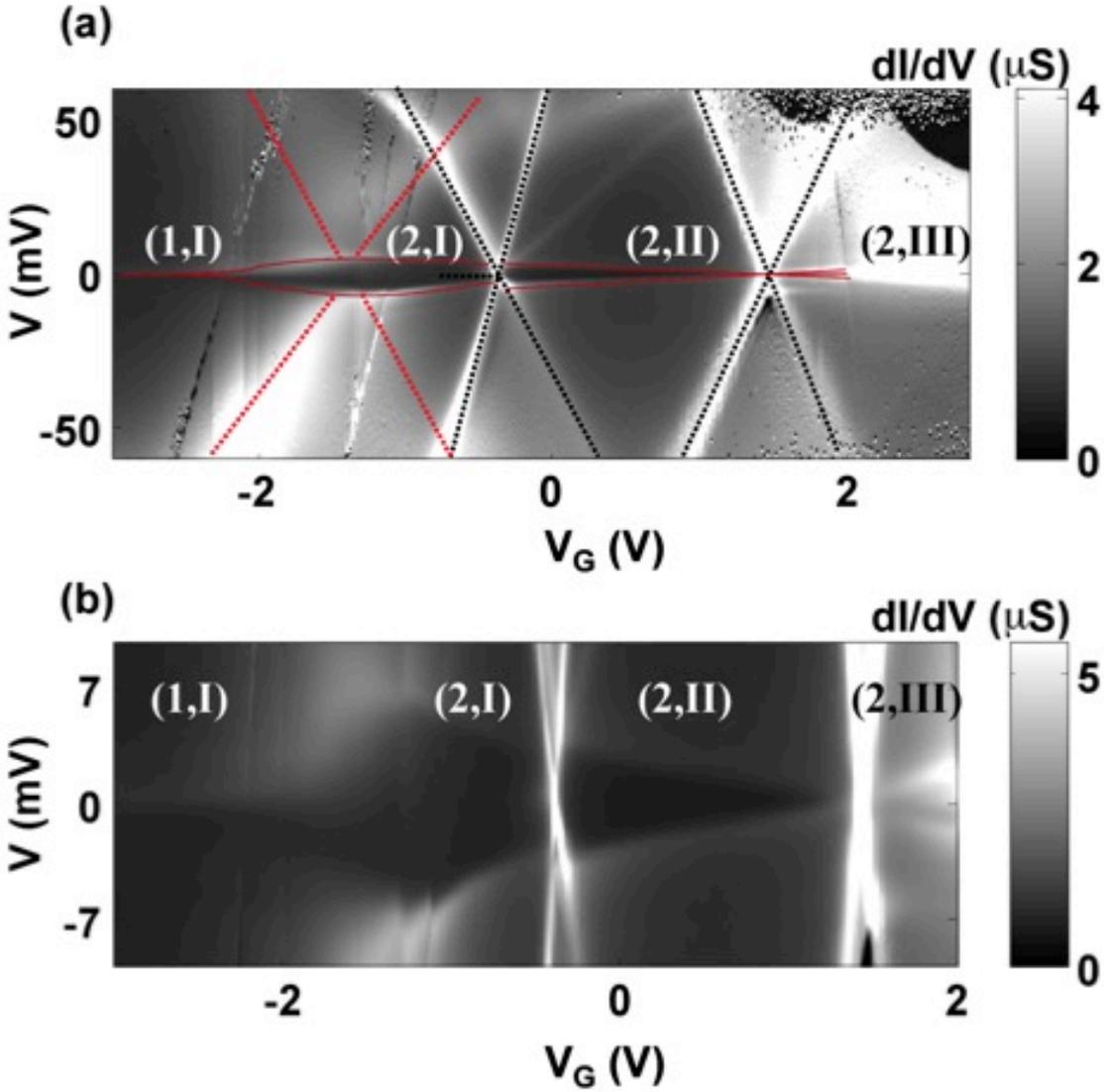

**Figure 2.** (a) Density plot of the differential conductance, $dI/dV$, versus $V$ and $V_G$ at $T = 1.7$ K. The different charge states of respectively the main molecule and the molecule in parallel are indicated by (i,j) with i=1,2 and j=I, II, III. The molecule in parallel gives rise to two very similar white crosses of high conductance due to sequential tunnelling (black dotted lines). The main molecule displays only a single cross (red dotted lines) corresponding to sequential tunnelling, which is strongly perturbed due to its very strong coupling to one, but not the other electrode, and due to the spin-blockade hindering ground state to ground state transport at low-bias. Solid red lines trace out the inelastic cotunneling edges due to virtual tunnelling processes in and out of charge state (2). Black areas at the top right of the figure are due to saturation of the current amplifier. (b) Low bias zoom-in of the different crossings and charge states without any guides to the eye.

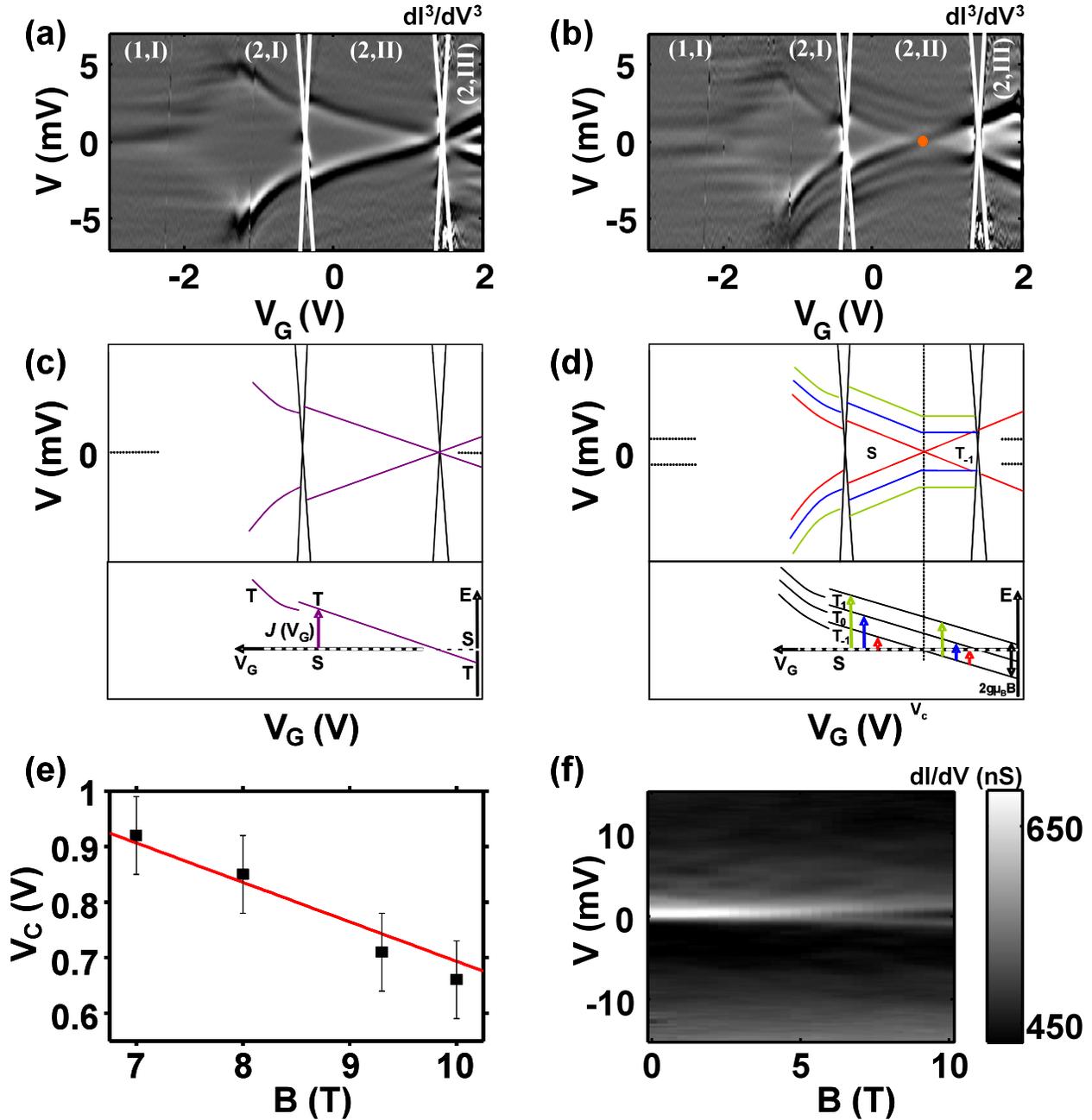

**Figure 3.** (a) Gray scale plot of $dI^3/dV^3$ as a function of $V$ and $V_G$ at $T$=1.7K and zero magnetic field obtained by numerical differentiation of the $dI/dV$ which was measured with a lock-in technique. We have plotted the third derivative in order to enhance the contrast of the low-bias features; resonances in the first derivative appear as dips (dark lines) in the third derivative. Almost vertical thick white lines

were superimposed on the plot at the diamond edge locations as a guide to the eye. (b) Same as (a) but at $B=10$T. (c) Top: schematic drawing of the important low bias features observed in (a). Bottom: energy diagram of the gate-dependent singlet (S) to triplet (T) transition observed in charge state (2). The energy splitting between S and T is given by the gate-dependent exchange coupling $J(V_G)$. (d) Top: schematic drawing of (b). Bottom: energy diagram with Zeeman splitting of the triplet states. Arrows indicate all observed transitions in charge state (2,I-III). The vertical dashed line locates the S-$T_{-1}$ crossing; as indicated on the top part, the singlet (triplet) is the ground state at the left (right) side of this line. (e) Squares: gate voltage value of the S-$T_{-1}$ crossing, $V_C$, for four different magnetic fields. Red line gives the predicted gate voltages of the S-$T_{-1}$ crossing, using the energy diagram presented in (d), as a function of magnetic field. (f) Gray scale plot of $dI/dV$ vs. $B$-field showing a Zeeman splitting of the Kondo resonance at $V_g = -2.8$V. From the splitting at $B=10$T we estimate a $g$-factor of $g = 1.9 \pm 0.3$.

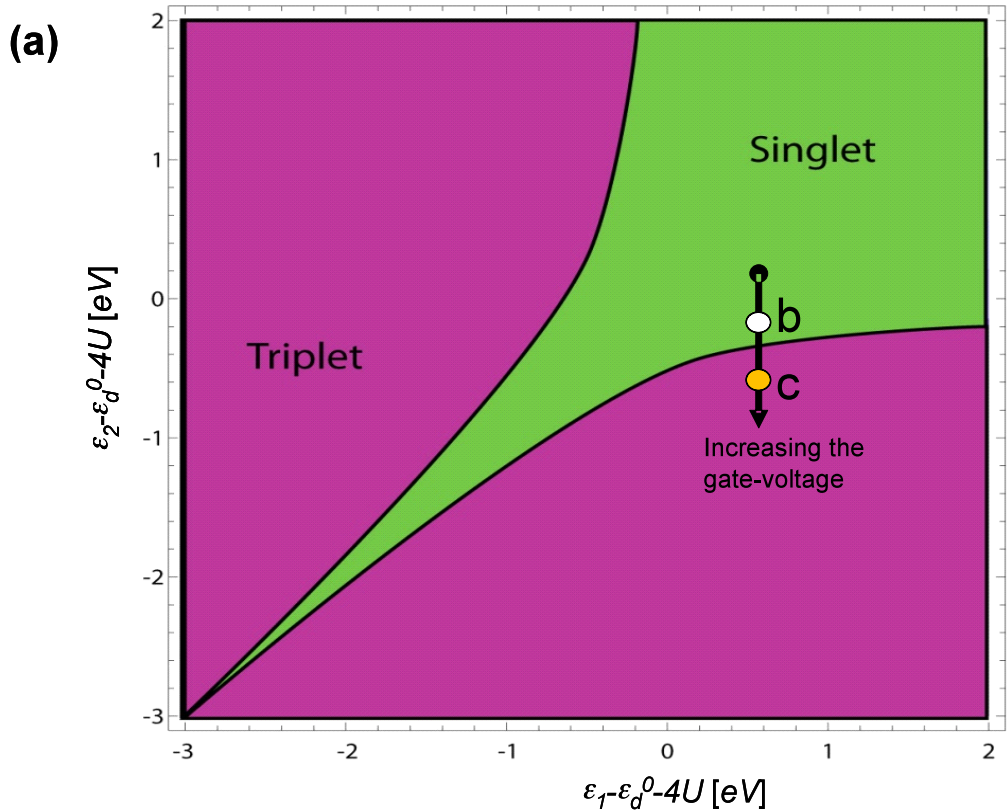

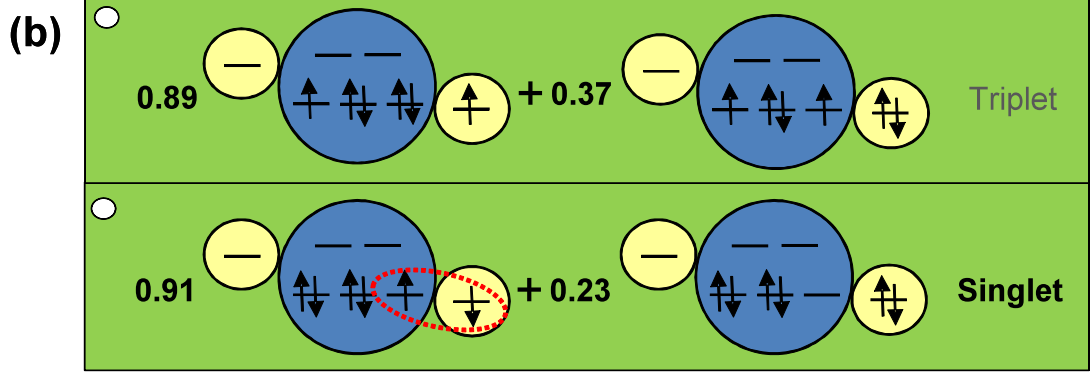

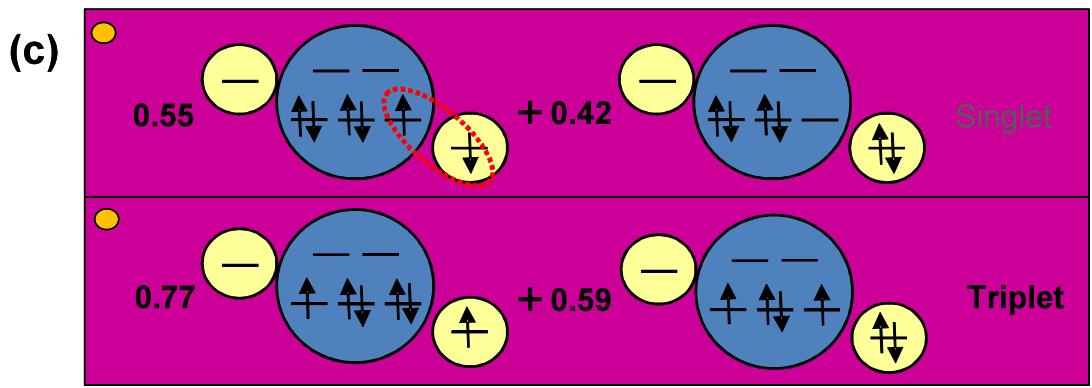

**Figure 4.** (a) Phase-diagram demarcating the regions for which the ground state of the *N*=6 electron system is respectively a spin singlet (green) or a spin triplet (purple). The control parameters on the axes represent the energy-levels of the relevant ligand states which hybridize with the *Mn d*-electrons. Due to the strongly asymmetric device geometry, all but the terpyridine moiety lying away from the leads will be screened by the nearby metallic lead and therefore the ligand level of this central terpyridine (ligand 2) feels the gate potential more strongly. Increasing *Vg* therefore lowers the energy of ligand 2 ($\varepsilon_2$ following the black arrow, say) and the ground state eventually changes from singlet to triplet as observed when moving away from the diagonal ($\varepsilon_1=\varepsilon_2$) in Fig. 3a and 3b. (b) The phase diagram in (a) is calculated from an exact diagonalization which reveals a simple understanding in terms of the 6-particle states shown here. Upper (lower) state is a spin singlet (triplet) and the singlet is the ground state at the point in parameter space corresponding to the white dot in panel (a). (c) Same as (b), except that $\varepsilon_2$ has now been moved down to the location of the orange dot in panel (a) and the triplet has become the ground state. As indicated by their numerical coefficients, these states dominate the exact eigenstates and they allow for a simple interpretation of the cause of the gate-dependent singlet-triplet splitting: Basically the triplet is stabilized by charge fluctuations between the *d*, and the ligand orbitals since it gains more from the Hund's rule coupling on the *Mn*-core (cf. section S6 of the supplementary information for more details). Increasing the gate-voltage lowers $\varepsilon_2$ and, as reflected in the different numerical coefficients, more weight is put on the component with a doubly charged ligand 2, i.e. the component where the triplet is lowered more in energy from Hund's rule coupling. Parameters for this plot were chosen to be *U*=5.0, $\varepsilon_d^0+4U=0$, *K*=0.8, *Δ*=2.0, *t*=0.26 and *t'*=0.1, all in units of eV.